\documentclass[pra,aps,twocolumn,showpacs]{revtex4-2}
\usepackage{amsmath,amssymb,graphicx}
\usepackage{float}
\usepackage{comment}
\usepackage{bm}
\usepackage{bbold}

\usepackage{graphicx}
\usepackage{dcolumn}
\usepackage{bm}
\usepackage{hyperref}
\usepackage{xcolor}
\usepackage{braket}
\usepackage{textcomp}
\usepackage{comment}
\usepackage{inputenc}

\DeclareMathOperator{\tr}{Tr}

\begin{document}

\preprint{APS/123-QED}

\title{Variational waveguide QED simulators}

\author{C. Tabares}
\affiliation{%
Institute of Fundamental Physics IFF-CSIC, Calle Serrano 113b, 28006 Madrid, Spain}%
\author{A. Muñoz de las Heras}
\affiliation{%
Institute of Fundamental Physics IFF-CSIC, Calle Serrano 113b, 28006 Madrid, Spain}%
\author{L. Tagliacozzo}
\affiliation{%
Institute of Fundamental Physics IFF-CSIC, Calle Serrano 113b, 28006 Madrid, Spain}%
\author{D. Porras}
\affiliation{%
Institute of Fundamental Physics IFF-CSIC, Calle Serrano 113b, 28006 Madrid, Spain}%
\author{A. González-Tudela}
\email{a.gonzalez.tudela@csic.es}
\affiliation{%
Institute of Fundamental Physics IFF-CSIC, Calle Serrano 113b, 28006 Madrid, Spain}%

\date{\today}

\begin{abstract}
Waveguide QED simulators are analogue quantum simulators made by quantum emitters interacting with one-dimensional photonic band-gap materials. One of their remarkable features is that they can be used to engineer tunable-range emitter interactions. Here, we demonstrate how these interactions can be a resource to develop more efficient variational quantum algorithms for certain problems. In particular, we illustrate their power in creating wavefunction ans\"atze that capture accurately the ground state of quantum critical spin models (XXZ and Ising) with less gates and optimization parameters than other variational ans\"atze based on nearest-neighbor or infinite-range entangling gates. Finally, we study the potential advantages of these waveguide ans\"atze in the presence of noise. Overall, these results evidence the potential of using the interaction range as a variational parameter and place waveguide QED simulators as a promising platform for variational quantum algorithms.
\end{abstract}

\maketitle

\emph{Introduction.-} Variational quantum algorithms (VQAs)~\cite{Cerezo2021,Bharti2022} aim at exploiting current noisy-intermediate scale quantum
 (NISQ) devices~\cite{preskill18a} before the fault-tolerant era arrives. Such algorithms leverage the power of classical optimizers to find the combination of single and multi-qubit gates (i.e., constructing a parametrized quantum circuit, or \emph{ansatz}) that miminizes a given cost function. The cost function is generally the expectation value of an operator in a state constructed using the parameterized quantum circuit, and measured in the quantum hardware. One paradigmatic example is the variational quantum eigensolver (VQE)~\cite{Tilly2022} in which the cost function is the energy of a given many-body Hamiltonian, e.g., in quantum chemistry~\cite{peruzzo14a,kandala17a,Quantum2020Hartree-FockComputer,Lee2019a, Matsuzawa2020,kivlichan17a,Setia2019} or high-energy physics problems~\cite{Kokail2019,Yamamoto2021,Irmejs2022,Paulson2021,Ciavarella2022}, among others. However, by changing the cost function, VQAs can also solve combinatorial optimization problems~\cite{Farhi2014,Lin2016,Moll2018,Wang2018,Lacroix2020ImprovingSets,Harrigan2021QuantumProcessor}, and be applied to quantum machine learning~\cite{Havlicek2019SupervisedSpaces,Johri2021NearestComputer} or quantum metrology protocols~\cite{Koczor2020,Kaubruegger2021,Ma2021,Beckey2022}.

 \begin{figure}[tb]
    \centering
    \includegraphics[width=\linewidth]{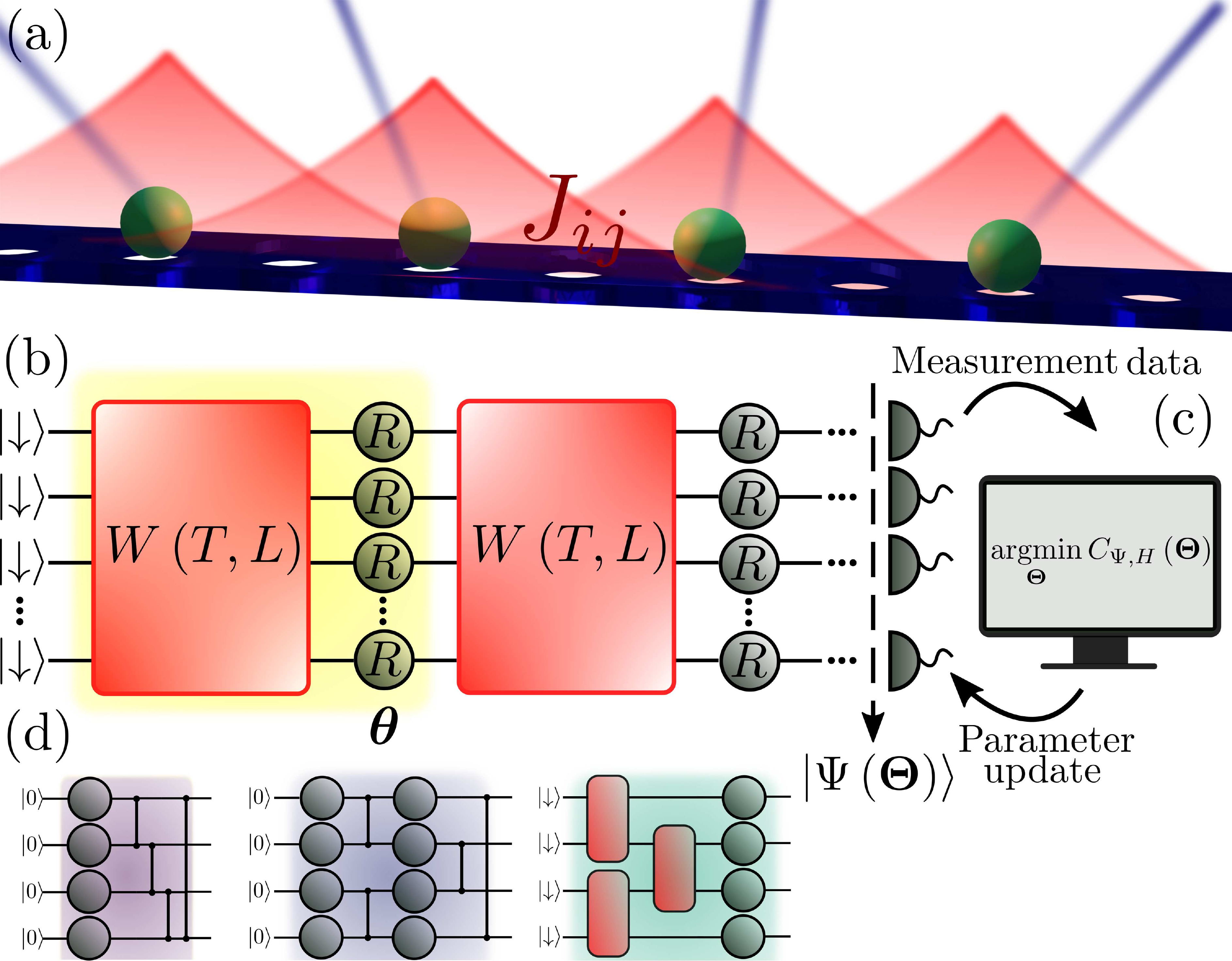}
      \caption{(a) Atoms (green) with optical transitions in waveguide band-gap regions form photonic bound-states (red), leading to tunable-range emitter interactions $J_{ij}\sim e^{-\left|x_i-x_j\right|/L}$ that can be controlled using external lasers (blue). (b) Taking the atoms as qubits, these interactions generate an entangling operation (red) which can be combined with single-qubit rotations $R\left(\bm{\theta}\right)$ to build a layer (yellow) of a variational quantum circuit that generates the state $\ket{\Psi\left(\bm{\Theta}\right)}$. (c) The measurement of an observable $H$ over this state defines a cost function $C_{\Psi,H}\left(\bm{\Theta}\right)$ that is minimized using a classical processor. (d) Scheme of other relevant circuits with first-neighbor entangling gates, such as the Hardware Efficient Ansatz (HEA) (purple), the brick-layer ansatz (dark blue), and the Hamiltonian Variational Ansatz (HVA) (light blue).}
    \label{fig:1}
\end{figure}

Like in classical variational approaches, the power of a VQA depends crucially on the ansatz. First, the ansatz needs to be expressive enough to accurately capture the solution of the problem targeted. While adding more gates allows to cover a wider Hilbert space region, circuits featuring a small number of them offer both a reduced complexity when estimating cost functions and their gradients through measurements~\cite{Tilly2022}, and are less prone to errors. Thus, an ansatz should ideally reproduce the phenomena of interest with as few gates as possible. Besides, it should avoid the \emph{barren-plateau problem}, i.e., the flattening of the optimization landscape, especially critical as the number of qubits increases~\cite{McClean2018} and for highly expressive ans\"atze~\cite{Holmes2022}. State-of-the-art ans\"atze divide between \emph{hardware-efficient ans\"atze}~\cite{kandala17a}, motivated by the connectivities of the devices, and \emph{problem-tailored} ones, such as unitary-coupled cluster~\cite{peruzzo14a} or Hamiltonian variational ans\"atze (HVAs)~\cite{wecker15a,Reiner2019,Verdon2019,Mele2022AvoidingAnsatz,Wiersema2020}, which are inspired by the problem structure. While the former are more naturally implemented in state-of-the-art NISQ devices, the latter are easier to optimize because their structure allows to avoid barren plateaus while being expressive enough to capture the solution~\cite{Wiersema2020}. However, such problem-inspired VQAs are also limited by the current hardware connectivities. Thus, despite many proof-of-principle VQA illustrations~\cite{peruzzo14a,kandala17a,Pagano2020QuantumSimulator,Kokail2019}, none of the ans\"atze considered fully solved all existing challenges. This is why the search for more efficient ans\"atze is one of the most pressing questions in the NISQ era.

In this work, we introduce and characterize a different type of ansatz inspired by the interactions that can be obtained in structured waveguide QED (wQED) setups~\cite{goban13a,goban15a,hood16a,Samutpraphoot2020,laucht12a,evans18a,Appel2021,Tiranov2022,MacHielse2019,Rugar2020,Rugar2021,liu17a,Mirhosseini2018a,Sundaresan2019,Scigliuzzo2022,Zhang2022Simulator,krinner18a}. These are systems where quantum emitters interact with one-dimensional photonic modes with non-linear energy dispersions [see Fig.~\ref{fig:1}(a)]. When the emitters' optical transition frequency lies within a photonic band-gap, the waveguide modes induce coherent, tunable-range interactions~\cite{douglas15a,Gonzalez-tudela2015b,Hung2016} which can be used to engineer multi-qubit entangling gates between the emitters. Here, we build ans\"atze combining such tunable-range gates with single qubit rotations [see Fig.~\ref{fig:1}(b)] and implement the VQE to show that they can represent the ground states (GSs) of critical spin models with fewer gates than existing ans\"atze. The key difference between the wQED ansatz and existing ones is the possibility to dynamically tune the interaction range, which we take as a variational parameter.  This allows the algorithm to find states displaying long-range correlations with less gates than the standard fixed-range connectivities ans\"atze. Finally, we study the impact of the reduced number of gates in the presence noise.

\emph{WQED ansatz.-} The physical setup we consider is summarized in Fig.~\ref{fig:1}(a). It consists of several emitters, that can be individually or globally addressed by lasers, coupled to a one-dimensional waveguide. Although we depict the emitters as atoms and the waveguide as a photonic-crystal one~\cite{Joannopoulos2011,Chang2018}, our findings can be extrapolated to other wQED platforms with different emitters (e.g., solid-state ones~\cite{laucht12a,evans18a,Appel2021,Tiranov2022,MacHielse2019,Rugar2020,Rugar2021}) and/or waveguides (such as microwave metamaterials~\cite{liu17a,Mirhosseini2018a,Sundaresan2019,Scigliuzzo2022,Zhang2022Simulator} or matter-wave ones~\cite{krinner18a}). Assuming the emitters couple to the waveguide modes through an effective transition between two states $g$ and $e$ whose frequency lies in the band-gap, and under the conditions in which the photonic field can be adiabatically eliminated, the dynamics of emitters is described by the following Hamiltonian~\cite{Gonzalez-tudela2015b,douglas15a,Hung2016}:
\begin{equation}\label{eq:H_XX}
    H_{\text{XX}} = J \sum_{i\neq j} e^{-|x_i-x_j|/L}\sigma^i_{\rm{eg}} \sigma^{j}_{\rm{ge}}\, ,
\end{equation}
where $J$ is the interaction strength, $L$ its effective range, which depends on the detuning between the effective transition frequency and the band-edge and thus can be dynamically tuned~\cite{bykov75a,john90a,kurizki90a}, $x_i$ is the atomic position, and $\sigma_{\rm ge}^i=\ket{g}_i\bra{e}_i$ the transition dipole operator. Similarly, as shown in Refs.~\cite{Gonzalez-tudela2015b,douglas15a,Hung2016} and Sup. Material (SM)~\cite{SupMatvarwqed}, if the emitters have two optically excited states that couple to the waveguide modes, one can also obtain an effective Hamiltonian $H_{\text{I}} = J \sum_{i\neq j} e^{-|x_i-x_j|/L}\sigma^i_{\rm{x}} \sigma^{j}_{\rm{x}}$, with $\sigma_{\rm x}^i=(\sigma^i_{\rm eg}+\sigma^i_{\rm ge})/2$ the Pauli matrix in the x-direction. Applying these Hamiltonians for a time $t$, one can obtain multi-qubit gates described by the unitaries
$W_{\rm XX}\left(T,L\right) = e^{-i t H_{\mathrm{XX}}}$ or $W_{\rm I}\left(T,L\right) = e^{-i t H_{\mathrm{I}}}$, respectively. Such gates can be parametrized by two tunable parameters [see Fig.~\ref{fig:1}(b)]: the normalized interaction time $T=t J$, and its range $L$. To build our wQED ans\"atze, we concatenate these unitaries with single-qubit rotations, e.g., in the $z$- direction, described by the unitary $R(\boldsymbol\theta)=\prod_{j=1}^N e^{-i\theta_{j} \sigma_{\mathrm{z}}^j}$, where $\boldsymbol\theta=(\theta_1,\theta_2,\dots)$. The wQED ansatz consists of $D$ repetitions or \emph{layers} of this combination of single and multi-qubit unitaries, see Fig.~\ref{fig:1}(b), described by the global unitary:
\begin{equation}\label{eq:wQED_ansatz}
U_{\mathrm{wQED-\alpha}}(\bm{\Theta}) = \prod_{i=1}^{D} R_{i}(\boldsymbol\theta^{i}) W_{i,\mathrm{\alpha}}(T_i,L_i)\,,
\end{equation}
where $\bm{\Theta}=(\boldsymbol\theta,\bm{T} , \bm{L} )$ is a vector embedding all variational parameters, and $\alpha=\mathrm{XX}$ or $\mathrm{I}$ depending on the coupling configuration chosen.

\begin{figure*}[tb]
    \centering
    \includegraphics[width=\linewidth]{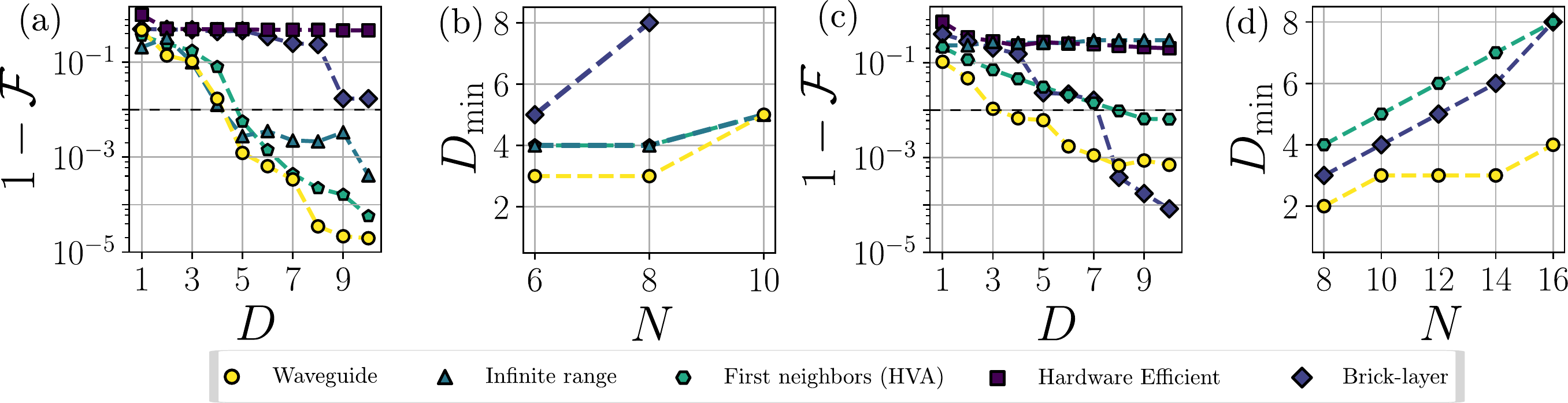}
    \caption{(a) Infidelity $1-\mathcal{F}$ between the exact GS of the XXZ model and the optimized variational states obtained with different ans\"atze (see legend below) as a function of the number of layers $D$ for a system with $N=10$ qubits. (b) Minimum depth $D$ required to obtain a fidelity $\mathcal{F}$ over 99\% for different numbers of qubits $N$ (note that not all the ansätze reach this fidelity). (c-d) Equivalent to (a-b) but considering the TFIM model instead of the XXZ model. The system size in (c) is $N=16$ qubits.}
    \label{fig:2}
\end{figure*}

\emph{VQE with wQED ans\"atze-} The main steps of VQE are~\cite{Tilly2022} (further explanations can be found in the SM~\cite{SupMatvarwqed}): i) The multi-qubit system is initialized in a state $\ket{\Psi_\mathrm{0}}$. ii) The initial state is the input of a parametrized quantum-circuit ansatz described by a unitary $U(\bm{\Theta})$, for some initial choice of variational parameters $\bm{\Theta}$. The output is a final state $\ket{\Psi(\bm{\Theta})}=U(\bm{\Theta})\ket{\Psi_\mathrm{0}}$. iii) The cost-function, i.e., the expectation value of the many-body Hamiltonian $C(\bm{\Theta})=\bra{\Psi(\bm{\Theta})}H\ket{\Psi(\bm{\Theta})}$ is obtained. iv) Finally, the value of $C(\bm{\Theta})$ is fed to a classical optimizer that updates the parameters. This procedure is repeated until the energy does not change significantly, meaning that VQE has found the optimal parameters $\bm{\Theta}_{\mathrm{opt}}$ that minimize the energy of $H$, unless it gets stuck in some local minima~\cite{McClean2018,Holmes2022}. Thus, this value of the energy is an upper bound to the actual value of the GS energy.

In what follows, we apply this procedure to several quantum critical spin models. These constitute interesting benchmarks as their GSs feature long-range correlations, which are the most challenging to capture for classical and quantum algorithms~\cite{Bravo-Prieto2020ScalingSystems,Jobst2022,Roca-Jerat2023}. We use an adiabatically-assisted VQE algorithm~\cite{Garcia-Saez2018AddressingEigensolvers}, explained in detail in SM~\cite{SupMatvarwqed}, and compare the performance of the wQED ans\"atze against the most popular fixed-structure ans\"atze of the literature [see Fig.~\ref{fig:1}(d) for a schematic picture of their circuits]: a hardware-efficient ansatz (HEA)~\cite{kandala17a} based on concatenating single qubit rotations along the three spatial directions, and two-qubit control-Z (CZ) as entangling gates; a brick-layer ansatz, in which the entangling CZ gates are introduced sequentially and interleaved by single qubit rotations; a first-neighbor HVA, in which the layers are given by the second-order trotterization of the spin Hamiltonian under consideration~\cite{Wiersema2020}; and also against an all-to-all ansatz, which is a wQED ansatz in which we take $L\rightarrow \infty$. The latter allows to discern whether a possible advantage is a consequence of the long-range character of the wQED interaction or its dynamical tunability. Besides, its results can be of interest for cavity QED setups~\cite{ritsch13a,Davis2019,Periwal2021,Li2022,Ramette2022,Greve2022,Hosten2016}, where such infinite range interactions appear naturally.

We first consider the XXZ model~\cite{Langari1998}
\begin{equation}\label{eq:HXZZ}
    H_{\mathrm{XXZ}} =\sum_{i} \left(\sigma_{\rm x}^i\sigma_{\rm x}^{i+1}+\sigma_{\rm y}^i\sigma_{\rm y}^{i+1}\right)-\Delta\sum_{i}\sigma_{\rm z}^{i}\sigma_{\rm z}^{i+1}
\end{equation}
at its ferromagnetic Heisenberg point $\Delta=1$. In Fig.~\ref{fig:2}(a-b) we evaluate the performance of the wQED-XX ansatz of Eq.~\ref{eq:H_XX} (in yellow circles) against the aforementioned ans\"atze. This wqED ansatz choice is owed to the resemblance of $H_{\rm XX}$ and the interactions appearing in the XXZ model. To assess the ans\"atze performance we compute the infidelity:
\begin{equation}
1-\mathcal{F}  =1- |\braket{\Psi(\bm{\Theta}_{\rm opt})|\Psi_\mathrm{GS}}|,
\label{eq:inf}
\end{equation}
with $\ket{\Psi_\mathrm{GS}}$ the GS  of the model, and $\ket{\Psi(\bm{\Theta}_{\rm opt})}$ that with the lowest energy found by the VQE. In Fig.~\ref{fig:2}(a) we plot the infidelity for $N=10$  qubits as a function of the circuit depth $D$ (i.e., the number of layers). Adding more layers typically results in better fidelities for most ans\"atze. This is expected since deeper circuits contain more variational parameters that can help exploring larger regions of the Hilbert space and getting closer to the GSs~\cite{Larocca2021TheoryNetworks,Bravo-Prieto2020ScalingSystems}. However, we appreciate that the wQED-XX ansatz performs better than the rest. To confirm this fact, we calculate the required circuit depth to obtain fidelities beyond 99\% for several $N$'s, finding that the wQED-XX ansatz produces a better approximation to the GS with shallower circuits than the other ans\"atze [see Fig.~\ref{fig:2}(b)]. This shallowness is advantageous because it makes the system more resilient to noise, and will require fewer measurements in the optimization loop.

This advantage is clearer when we study the transverse-field Ising model (TFIM)
\begin{equation}
    \label{eq:HIsingmain}
    H_{\mathrm{Ising}}=-\sum_{i}\sigma_{\rm x}^i\sigma_{\rm x}^{i+1}+g\sum_i \sigma^i_{\rm z}\,,
\end{equation}
at the critical point $g=1$. The results are shown in Fig.~\ref{fig:2}(c-d). In this case we use the wQED-I ansatz, again due to the similarity between the Ising interactions appearing in $H_\mathrm{I}$ and those of the target problem. On top of that, it is noteworthy that here we use a global single-qubit rotation, which substantially reduces the number of parameters with respect to the other ans\"atze~\cite{SupMatvarwqed}, and simplifies the experimental implementation because one can address the emitters globally. The results of the ans\"atze comparison are presented in Fig.~\ref{fig:2}(c). There, we fix the number of qubits at $N=16$, and show that for most ans\"atze  the infidelity decreases with the circuit depth $D$. Due to the particular structure of the wQED-I ansatz, in this case we can benchmark the ans\"atze for larger values of $6\le N\le 16$, for which it provides the most accurate results, especially at low circuit depths. This is confirmed in Fig.~\ref{fig:2}(d), where we plot the depth required by the ans\"atze to achieve fidelities beyond 99\%. There, the wQED-I ansatz reaches the desired fidelity systematically for smaller $D$. It is also interesting to highlight the inferior performance of the all-to-all ansatz compared with both wQED ans\"atze. This points to the fact that it is the dynamical tuning of the interaction range rather than its long-range character what allows the latter to reproduce better the power-law correlations of the GSs, and thus to outperform the rest. 

\begin{figure}[tb]
    \centering
    \includegraphics[width=\linewidth]{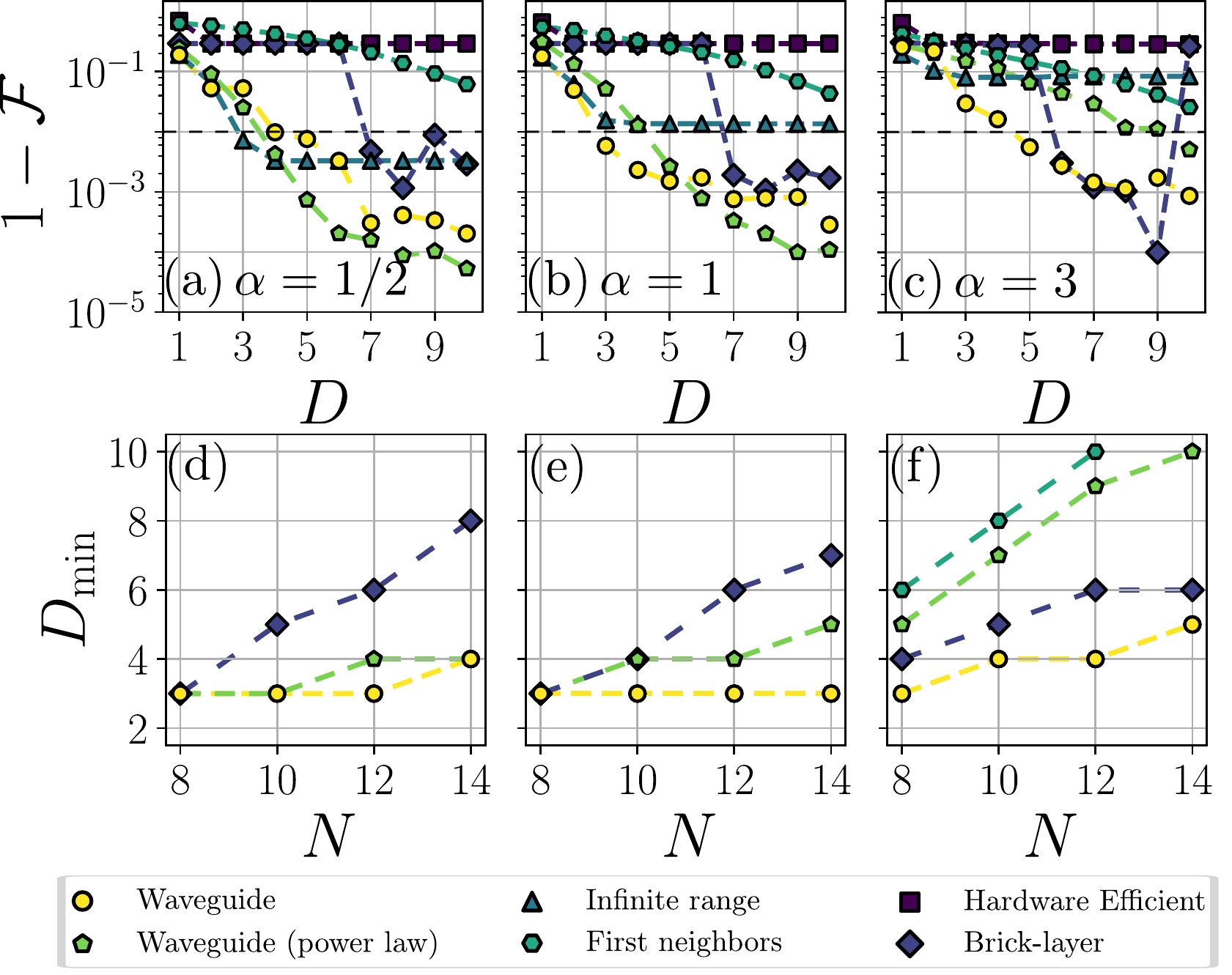}
    \caption{(a-c) Infidelity $1-\mathcal{F}$ between the exact GS of the LRTFIM and the optimized variational states obtained with different ans\"atze as a function of the number of layers $D$ for several values of the power-law exponent $\alpha$. The system size is $N=14$ qubits. (d-f) Minimum depth $D$ required to obtain a fidelity $\mathcal{F}$ over 99\%  as a function of the number of qubits $N$ and for different values of $\alpha$.}
    \label{fig:3}
\end{figure}

Finally, we check the performance of the wQED ansatz for models with long-range interactions, like the long-range transverse-field Ising model (LRTFIM):
\begin{equation}\label{eq:H_LRTFIM}
    H_{\text{LR}} = -\sin{\theta}\sum_{i\neq j} \frac{1}{|i-j|^{\alpha}} \sigma_{\rm x}^{i}\sigma_{\rm x}^{j} + \cos{\theta}\sum_{i=1}^{N}  \sigma_{\rm z}^{i}\,.
\end{equation}
In this case the position of the critical point depends on the power-law exponent $\alpha$~\cite{Ruelle1968,Dyson1969,Dutta2001}: here we consider three different values $\alpha=1/2,1,3$, and choose $\theta$ to be at the critical point~\cite{SupMatvarwqed} in each case. The interactions appearing in the target Hamiltonian~\eqref{eq:H_LRTFIM} make the wQED-I ansatz of Eq.~\eqref{eq:wQED_ansatz} the natural choice for this problem. Besides, we also compare with a modified version of the wQED-I ansatz in which $W_{\rm I}(T_i,L_i)$ is replaced by the product of two unitaries $\prod_{l=1}^2 W_{\rm I}(T_l,L_l)$ within each layer, and with the parameters $T_l,L_l$ chosen in such a way that $\sum_l J_l e^{-|x_i-x_j|/L_l}\approx J_\alpha/|x_i-x_j|^\alpha$ approximates the power-law exponent of the target model for a given range of distances~\footnote{These parameters can always by found by  fitting the power-law with exponentials~\cite{Pirvu2010MatrixRepresentations}}. This multi-exponential interaction can be realized by means of multi-frequency Raman lasers, as shown in Ref.~\cite{douglas15a}. The results are summarized in Fig.~\ref{fig:3}, where the panels (a-c) display the infidelity as a function of the number of layers for $N=14$ qubits, and (d-f) show the required depth to achieve a fidelity beyond $99\%$ for different $N$'s. The main conclusion is that, like in the previous cases, the wQED-I ans\"atze achieves the targeted fidelities for smaller (or equal) circuit depths, even for the longer-ranged models. However, in these models the power-law wQED ansatz eventually outperforms the exponential one for larger circuit depths.

\begin{figure}[tb]
    \centering
    \includegraphics[width=0.9\linewidth]{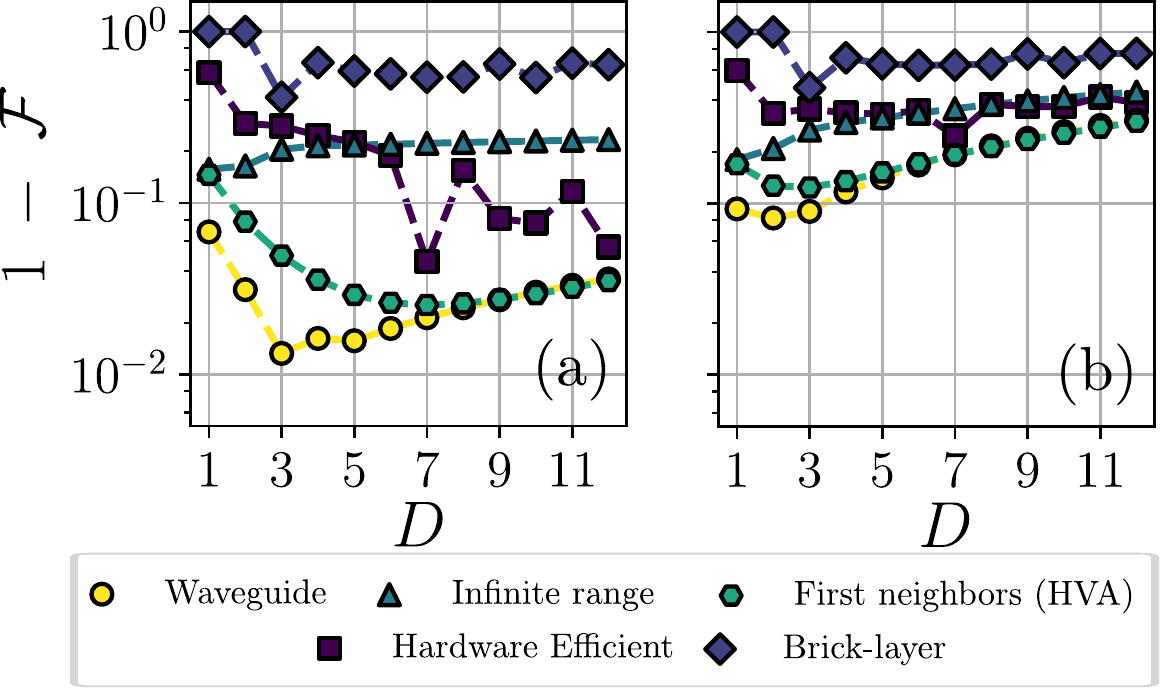}
    \caption{(a-b) Effects of decoherence in the infidelity $1-\mathcal{F}$ between the exact GS of the TFIM and the optimized variational states obtained with different ans\"atze as a function of the number of layers $D$ for $N=12$ qubits. The error probabilities for single and two qubit gates are, respectively: $p_1 = 10^{-5}$ and $p_2 = 5\times10^{-4}$ in (a), and $p_1 = 10^{-4}$ and $p_2 = 5\times 10^{-3}$ in (b).
   }
    \label{fig:4}
\end{figure}

\emph{Impact of noise.-} So far we have considered an ideal noiseless situation. This is not the case of real setups where dissipation associated to the waveguide modes, disorder, and other decay channels will introduce errors. Given the variety of wQED platforms~\cite{goban13a,goban15a,hood16a,Samutpraphoot2020,laucht12a,evans18a,Appel2021,Tiranov2022,MacHielse2019,Rugar2020,Rugar2021,liu17a,Mirhosseini2018a,Sundaresan2019,Scigliuzzo2022,Zhang2022Simulator,krinner18a}, where the most relevant errors might have a different physical origin, to study the impact of noise we use the platform-agnostic error model described in Refs.~\cite{kandala17a,SupMatvarwqed}, which assumes a constant error probability $p_{1 (2)}$ for each single (multi)-qubit gate. We focus on the impact of noise in the VQE for the TFIM in Eq.~\eqref{eq:HIsingmain} as cost function, since it is numerically easier. The results are summarized in Fig.~\ref{fig:4}, where we show the evolution of the infidelity with the number of layers for two values of $p_{1 (2)}$, each pair corresponding to a different panel. Such values have been chosen in the range of gate fidelities displayed by some of the state-of-the-art NISQ devices~\cite{IonQ,IBMQuantum}. We conclude that in the presence of noise the infidelity does not decrease monotonically when increasing the circuit depth. This is true for all ans\"atze, including the wQED-I one. The reason behind is that by adding more layers one introduces more variational parameters, which improves the optimization process at equal conditions, but the resulting increase in accuracy is lost at some point due to the error accumulation by the application of more gates. With this error model, the wQED-I ansatz reaches the smallest infidelity. The underlying reason is a combination of its better performance at small circuit depths, and the smaller number of gates per layer with respect to the other ans\"atze, which limits the introduction of errors. The latter will be true also for errors coming from imperfect gate control, which could also benefit from the smaller depth of the wQED circuits. We leave for future work the study of more refined error models devoted to each specific implementation~\cite{goban13a,goban15a,hood16a,Samutpraphoot2020,laucht12a,evans18a,Appel2021,Tiranov2022,MacHielse2019,Rugar2020,Rugar2021,liu17a,Mirhosseini2018a,Sundaresan2019,Scigliuzzo2022,Zhang2022Simulator,krinner18a}.

\emph{Conclusions\& Outlook.-} To sum up, we introduce a new type of VQE ansatz based on the tunable-range interactions that can be engineered in wQED setups. We show that, thanks to this tunability, this ansatz captures phases with long-range correlations with less gates and parameters than other fixed structure ans\"atze of the literature, becoming potentially less sensitive to gate errors. While wQED is still behind in fidelities compared to variational trapped ions~\cite{Kokail2019,Kaubruegger2021} or superconducting quantum simulators~\cite{kandala17a}, the rapid experimental advances in the integration of emitters with photonic waveguides~\cite{goban13a,goban15a,hood16a,Samutpraphoot2020,laucht12a,evans18a,Appel2021,Tiranov2022,MacHielse2019,Rugar2020,Rugar2021}, microwave circuits~\cite{liu17a,Mirhosseini2018a,Sundaresan2019,Scigliuzzo2022, Zhang2022Simulator} or matter-waves~\cite{krinner18a}, plus the potentialities shown in this work, place wQED simulators as promising candidates for VQAs. Besides, our idea of using the interaction range as a variational parameter can be extended to other setups where such tunable-range interactions can be engineered such as trapped-ions~\cite{porras04a,nevado16a,richerme14a,Jurcevic2014, Joshi2020,Monroe2021} or multi-mode cavity QED setups~\cite{Vaidya2018}. As an outlook, we plan to extend our results to higher-dimensional models~\cite{Armon2021}, where such tunable exponential~\cite{Gonzalez-tudela2015b,gonzaleztudela18f} and power-law interactions can also be obtained~\cite{Perczel2020a,Gonzalez-Tudela2018,Redondo-Yuste2021a,Garcia-Elcano2020,Navarro-Baron2021a}, models with high-dimensional spins~\cite{Davis2019,Orioli2021,Tabares2022}, as well as to design adaptative ans\"atze~\cite{Zhu2022,Grimsley2019,Yao2021,Zhang2021,Claudino2020,Tang2021Qubit-ADAPT-VQE:Processor} based on wQED interactions, which is another active area of research within variational quantum computing.

\begin{acknowledgements}
  The authors acknowledge support from the Proyecto Sin\'ergico CAM 2020 Y2020/TCS-6545 (NanoQuCo-CM), the CSIC Research Platform on Quantum Technologies PTI-001 and from Spanish projects PID2021-127968NB-I00 and TED2021-130552B-C22 funded by r MCIN/AEI/10.13039/501100011033/FEDER, UE and MCIN/AEI/10.13039/501100011033, respectively. The authors also acknowledge Centro de Supercomputación de Galicia (CESGA) who provided access to the supercomputer FinisTerrae for performing numerical simulations. AGT also acknowledges support from a 2022 Leonardo Grant for Researchers and Cultural Creators, BBVA Foundation, and thanks X. Zhang for insightful discussions on the physical implementation of the ideas and for a critical reading of the manuscript.
\end{acknowledgements}

\bibliography{references_alex}

\newpage
\begin{widetext}

\begin{center}
\textbf{\large Supplementary Material: Variational waveguide QED simulators \\}
\end{center}
\setcounter{equation}{0}
\setcounter{figure}{0}
\makeatletter

\renewcommand{\thefigure}{SM\arabic{figure}}
\renewcommand{\thesection}{SM\arabic{section}}  
\renewcommand{\theequation}{SM\arabic{equation}}  

In this Supplementary Material, we provide more details that support the results presented in the main text. In Section~\ref{secSM:wQED} we review the physical origin of the tunability of the photon-mediated interactions when the waveguide modes feature band-gaps. Then, in Section~\ref{secSM:VQE} we summarize the key steps of the variational quantum eigensolver algorithm (\ref{subsecSM:VQEkey}), describe the properties of target models considered (\ref{subsecSM:models}), explain the structure of the different ans\"atze typically employed in the literature (\ref{subsecSM:ansatz}), give details about the optimization protocol we use to obtain the results of the main text (\ref{subsecSM:opt}) and comment about other possible figures of merit to benchmark our results (\ref{subsecSM:fig_of_merit}). Finally, in Section~\ref{secSM:error} we discuss the details of the error model we use to obtain Fig.~\ref{fig:3} of the main text. Let us also note that all the codes to reproduce the results of the manuscript are available at \url{https://github.com/cristiantlopez/Variational-Waveguide-QED-Simulators}.



\section{Photon-mediated interactions in waveguide QED with band-gaps~\label{secSM:wQED}}

The photon-mediated interactions in photonic-bandgap waveguides have been considered extensively in other works~\cite{douglas15a,Gonzalez-tudela2015b,Hung2016}. For this reason here we will just review how they emerge, aiming at providing the reader with an intuition of the origin of their tunability.

Let us consider a system like the one depicted in Fig.~\ref{fig:1}(a) of the main text featuring $N_e$ emitters interacting with a photonic-crystal waveguide with energy dispersion $\omega_k$. In that case, the light-matter interaction Hamiltonian is composed by three parts. First, the photonic part given by
\begin{align}
\label{eqSM:HB}
H_{\rm B}=\sum_k \omega_k a_k^\dagger a_k\,,
\end{align}
being $a_k^{(\dagger)}$ the destruction (creation) operator of a waveguide photon with momentum $k$. Second, the emitters' Hamiltonian, $H_{\rm S}$, which will depend on the particular level structure. For the moment we do not write an explicit form, and just assume is that they have at least one optical transition that couple to the waveguide mode through a general dipole operator $\mathcal{O}_i$ with an effective frequency $\omega_\mathcal{O}$ equal for all the emitters. In that case, the last term of the total Hamiltonian is the light-matter coupling Hamiltonian, that can be written
as
\begin{align}
\label{eqSM:HLM}
H_{\rm I}(t)=\sum_{k,i} g_{k}e^{-i k x_i} a_k^\dagger \mathcal{O}_i e^{i(\omega_k-\omega_\mathcal{O})t}+\mathrm{H.c}\,,
\end{align}
where $g_{k}$ is the single-photon coupling strength of the $i$-th emitter to the photon with $k$-momentum. Here, we implicitly assume that the coupling is local, so that the only phase appearing in the light-matter coupling is the propagating phase $e^{-i k x_i}$, and small compared to the optical transition frequencies so that we only keep those terms conserving the number of excitations. Both are good approximations for the implementations that we are interested in.

In this manuscript we are interested in the situation in which the photonic field can be eliminated and induces an effective dynamics on the emitters. Assuming that the coupling is weak and the photonic bath is memory-less (Born-Markov approximation), the effective emitter dynamics in the interaction picture is given by
\begin{equation}\label{eq:sup_waveguide_mastereq}
    \frac{\mathrm{d}\rho}{\mathrm{d}t} = -i\left[\sum_{i,j} J_{i,j}\mathcal{O}_i^\dagger \mathcal{O}_j,\rho\right] +\sum_{i,j}\frac{\gamma_{i,j}}{2}\left(2\mathcal{O}_j\rho\mathcal{O}^\dagger_i-\mathcal{O}_i\mathcal{O}_j^\dagger\rho-\rho\mathcal{O}_i\mathcal{O}_j^\dagger\right)\,,
\end{equation}
where the photon-mediated interactions read~\cite{douglas15a,Gonzalez-tudela2015b,Hung2016}
\begin{align}
J_{i,j}-i\frac{\gamma_{i,j}}{2}=\int \frac{\mathrm{d}k}{2\pi}\frac{|g_k|^2 e^{i k(x_i-x_j)}}{\omega_\mathcal{O}-\omega_k+i0^+}\,,\label{eqSM:dipole}
\end{align}
and depend on the detuning $\omega_{\mathcal{O}}-\omega_k$ which appears in the light-matter Hamiltonian~\eqref{eqSM:HLM}. From here we see that if $\omega_{\mathcal{O}}
$ lies in a photonic band-gap of $\omega_k$, i.e., $\omega_{\mathcal{O}}
\neq \omega_k$ for any $k$, the non-unitary part cancels, i.e., $\gamma_{i,j}\equiv 0$. Besides, if we assume that around the band-edge $\omega_k\approx \omega_c+A (k-k_c)^2$, then it can be shown that $J_{i,j}$ is exponentially localized
\begin{align}
J_{i,j}=J e^{-|x_i-x_j|/L}\,,
\end{align}
with a localization length $L=\sqrt{A/(\omega_c-\omega_\mathcal{O})}$ that depends on the relative detuning between the band-edge and the dipole operator's frequency, and $J\propto |g_{k_\mathcal{O}}|^2$. Here, we have also absorbed the relative phases $e^{i k_c |x_i-x_j|}$ which can be corrected by appropriately placing the emitters.
\begin{figure}[t]
    \centering
    \includegraphics[width=\linewidth]{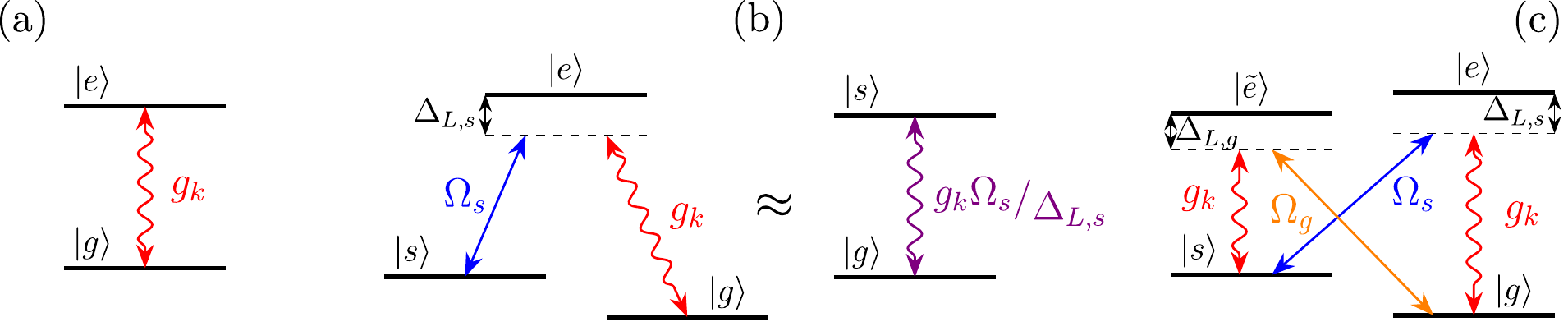}
    \caption{Sketches of the different emitters' energy levels, couplings to the photonic modes, and laser-driven transitions used to obtain the effective Hamiltonians discussed in the main text. (a) The transition between a ground and an excited state, $\ket{g}$ and $\ket{e}$ respectively, can be coupled to a photonic mode $k$ that, once adiabatically eliminated, lead to the effective Hamiltonian shown in Eq.~\ref{eqSM:HXX}. (b) If an additional level in the ground state manifold, $\ket{s}$, is coupled with amplitude $\Omega_s$ and detuning $\Delta_{L,s}$ to the excited state $\ket{e}$, and under the assumption that $\left|\Omega_s\right|\ll \Delta_{L,s}$ (i.e., a far detuned laser), the excited level can be adiabatically eliminated to find an effective light-mediated interaction between the emitters as the one shown in Fig.~\ref{fig:SM1_effective-Hamiltonians}(a) but whose amplitude can be dynamically tuned, c.f. Eq.~\ref{eqSM:raman_HXX}. (c) Furthermore, if an additional excited state $\ket{\tilde{e}}$ (coupled to the same photonic modes $a_k$ via the $\ket{\tilde{e}}\leftrightarrow\ket{s}$ transition) is coupled to the ground state $\ket{g}$ via a Raman laser of amplitude $\Omega_g$ and detuning $\Delta_{L,g}$, the adiabatic elimination of the excited states leads to an effective Ising-type interaction between the emitters, Eq.~\ref{eqSM:H_Ising}.}
    \label{fig:SM1_effective-Hamiltonians}
\end{figure}
Now, let us explain the different situations considered in the main text:
\begin{itemize}
\item First, the simplest instance when we have a two level emitter with a single ground ($g$) and excited ($e$) state, shown in Fig.~\ref{fig:SM1_effective-Hamiltonians}(a). In that case, the emitters' dipole operators are given by $\mathcal{O}_i=\sigma_{ge}^i$, and $\omega_\mathcal{O}=\omega_{e}-\omega_g$ is simply the energy difference between the ground and excited states. Thus, the effective interaction Hamiltonian induces spin-exchange excitations between the ground and excited state of the emitters:
\begin{align}
H_\mathrm{XX}=J\sum_{i,j}e^{-|x_i-x_j|/L}\sigma_{\rm eg}^i\sigma_{\rm ge}^j\,,\label{eqSM:HXX}
\end{align}
which is sometimes referred to in the literature as XX interaction. To be able to switch on and off the interaction $J$ one would need a mechanism to switch the coupling to the waveguide modes, whereas to tune $L$ one needs a dynamical control either of the waveguide/emitter frequencies.

\item In the case where the emitter has an additional ground state $s$ connected via a Raman transition with intensity $\Omega_s$ and frequency $\omega_{L,s}$ to the excited state $e$, as shown in Fig.~\ref{fig:SM1_effective-Hamiltonians}(b), one can further tune the dynamics. In particular, if the detuning of the Raman laser with respect to the optically excited state, $\Delta_{L,s}=\omega_e-\omega_{L,s}-\omega_s$ is very large compared to the intensity $\Omega_s$, the excited state is only virtually populated and thus one obtains a renormalized light-matter interaction Hamiltonian given by
\begin{align}
\label{eqSM:HLM1}
H_I\approx \sum_{k,i} g_{k}\frac{\Omega_s}{2\Delta_{L,s}}e^{-i k x_i} a_k^\dagger \sigma_{gs}^i e^{i\left(\omega_k-\omega_s-\omega_{L,s}+\omega_g\right)t}+\mathrm{H.c}\,,
\end{align}
in which the effective dipole operator is $\mathcal{O}_i=\sigma_{gs}^i$. Here, we observe two interesting consequences: On the one hand, the effective light-matter coupling is renormalized by $\Omega_s/2\Delta_{L,s}$, which enables to switch it on and off. On the other hand, the frequency of the laser appears in the interaction Hamiltonian, which enables to control the effective frequency of the interaction $\omega_\mathcal{O}=\omega_L+\omega_s-\omega_g$.  Thus, in this case the shape of the photon-mediated interactions is qualitatively similar to Eq.~\eqref{eqSM:HXX}:
\begin{align}\label{eqSM:raman_HXX}
H_{\mathrm{XX}}=\tilde{J}\sum_{i,j} e^{-|x_i-x_j|/\tilde{L}}\sigma_{sg}^i\sigma_{gs}^j\,,
\end{align}
but now the spin exchange occurs between the ground state levels $\{g,s\}$, and both $\tilde{J}$ and $\tilde{L}$ can be dynamically tuned.

\item If one considers that an additional excited state level ($\tilde{e}$) couples to the same waveguide modes $a_k$ via the additional state level $s$, as shown in Fig.~\ref{fig:SM1_effective-Hamiltonians}(c), one can  qualitatively change the shape of the interactions, as demonstrated in Refs.~\cite{douglas15a,Hung2016}. For that, one requires an additional Raman laser connecting the ground state $g$ with the additional excited state $\tilde{e}$, as depicted in Fig.~\ref{fig:1}(a) of the main text. Denoting by $\Omega_g$ and $\omega_{L,g}$ its amplitude and frequency, and assuming the same conditions as for the other excited state such that it can be adiabatically eliminated, one obtains the following renormalized light-matter interaction Hamiltonian:
\begin{align}
\label{eqSM:HLM2}
H_I\approx \sum_{k,i} g_{k}e^{-i k x_i} a_k^\dagger\left( \frac{\Omega_s}{2\Delta_{L,s}}\sigma_{gs}^i e^{i\left(\omega_k-\omega_s-\omega_{L,s}+\omega_g\right)t}+\frac{\Omega_g}{2\Delta_{L,g}}\sigma_{sg}^i e^{i\left(\omega_k+\omega_s-\omega_{L,g}-\omega_g\right)t}\right)+\mathrm{H.c}\,.
\end{align}

Fixing $2(\omega_g-\omega_s)=\omega_{L,s}-\omega_{L,g}$ and $\Omega_s/\Delta_{L,s}=\Omega_g/\Delta_{L,g}$, one obtains that the effective dipole operator that couples to the photonic bath is $\mathcal{O}_i=\sigma_{gs}^i+\sigma_{sg}^i=2\sigma_x^i$, i.e., the $x$-Pauli matrix, whereas $\omega_\mathcal{O}=\omega_s+\omega_{L,s}+\omega_g$. Thus, when adiabatically eliminating the photons one obtain an Ising-type interaction
\begin{align}\label{eqSM:H_Ising}
H_{\mathrm{Ising}}=J\sum_{i,j} e^{-|x_i-x_j|/L}\sigma_{x}^i\sigma_{x}^j\,,
\end{align}
with also tunable $J$ and $L$.

\item Finally, let us consider that the Raman lasers have not a single frequency but rather several sidebands, e.g., $\Omega_{g/s}e^{i\omega_{L,g/s}t}\rightarrow \Omega_{g/s}\sum_{\alpha}c_\alpha e^{i\delta_\alpha t}$. As shown in Refs.~\cite{douglas15a,Hung2016}, each of the sidebands will generate a bound-state with different localization length $L_\alpha$ and different strength $J_\alpha$, which will depend on the weight and frequency of the sideband $\{\delta_\alpha,c_\alpha\}$. Thus, we will have Hamiltonians of the form
\begin{align}
H_\mathrm{R}=\sum_{i,j}\left(\sum_{\alpha}J_\alpha e^{-|x_i-x_j|/L_\alpha}\right)\mathcal{O}^\dagger_i\mathcal{O}_j\,,
\end{align}
where the particular operator $\mathcal{O}_i$ will depend on the Raman configuration chosen. This multi-exponential behaviour is interesting because, as noted in Ref.~\cite{douglas15a}, one can obtain an effective power-law behavior by adding exponentials. In fact, there exist constructive algorithms~\cite{Pirvu2010MatrixRepresentations} that give you the required $\{J_\alpha,L_\alpha\}$ to mimic a power-law dependence:
\begin{align}
\sum_{\alpha}J_\alpha e^{-|x_i-x_j|/L_\alpha}\approx \frac{J}{|x_i-x_j|^\alpha}\,.
\end{align}

In the main text, we use one of these algorithms to mimic the power-law dependence of the interactions of the models studied in Fig.~\ref{fig:4}.

\end{itemize}

\section{Variational quantum eigensolvers with fixed structure ans\"atze~\label{secSM:VQE}} 

\subsection{Summary of the key steps of variational quantum eigensolvers~\label{subsecSM:VQEkey}}

VQAs encode a problem into a cost function whose minimum corresponds to the solution of the problem. Typically, cost functions are of the form
\begin{equation}
    C_{\rho,H}\left(\bm{\Theta}\right) = \tr\left[H U\left(\bm{\Theta}\right)\rho U\left(\bm{\Theta}\right)^\dagger\right]\,, 
\end{equation}
where $\rho$ is an input $n-$qubit state, $H$ is an Hermitian operator and $U\left(\bm{\Theta}\right)$ is a parametrized quantum circuit in terms of the trainable sets of parameters $\bm{\Theta}$. Both the value of the cost function $C_{\rho,H}\left(\bm{\Theta}\right)$ and its derivatives can be calculated using a quantum computer, and then be fed into a classical optimizer to solve the problem:
\begin{equation}
    \left(\bm{\Theta}_\text{opt}\right) = \mathrm{arg min}_{\bm{\Theta}} C_{\rho,H}\left(\bm{\Theta}\right)\,.
\end{equation}
When the operator $H$ is taken as the Hamiltonian of a particular system the algorithm described above is called Variational Quantum Eigensolver (VQE)~\cite{peruzzo14a}, and the set of optimal parameters $\bm{\Theta}_\text{opt}$ yields a variational approximation $\rho\left(\bm{\Theta}_\text{opt}\right) = U\left(\bm{\Theta}_\text{opt}\right)\rho U\left(\bm{\Theta}_\text{opt}\right)^\dagger $ of the ground state of $H$. Considering that the (ideal) output of the quantum circuit is a pure state $\ket{\Psi\left(\bm{\Theta}\right)}$, the cost function becomes:
\begin{equation}\label{eqSM:cost_fun_energy}
    C_{\Psi,H} \left(\bm{\Theta}\right) := \left\langle\Psi\left(\bm{\Theta}\right)|H|\Psi\left(\bm{\Theta}\right)\right\rangle\,,
\end{equation}
which corresponds to the energy of the variational state. 

The set of optimal parameters $\bm{\Theta}_{\text{opt}}$ provides a good approximation to the solution of the optimization task depending on several factors: first, it is necessary to choose an adequate $H$. In the context of VQEs, $H$ is simply the Hamiltonian whose ground state we are looking for, but multiple choices for $H$ can be made in the context of other algorithms, and therefore another factors should be considered (such as how easy the operator $H$ can be measured in the quantum computer). Second, it is necessary to choose an ansatz $U\left(\bm{\Theta}\right)$ expressive enough to produce a state close to the solution of the problem. Typically, these ans\"atze are written as
\begin{equation}\label{eq:general_ansatz}
    U\left(\bm{\Theta}\right) = \prod_{i=1}^{D} \prod_{j=1}^{N} R^{j}_{i}(\theta^{i}_{j}) W_i\left(\bm{\phi}_{i}\right)\,,
\end{equation}
where the circuit is composed of $D$ layers, each formed by a set of single-qubit rotation gates $R^{j}_{i}=e^{-i\theta_{i}^{j} V^j}$ (with $V^j \in \left\{\sigma^x,\sigma^y,\sigma^z\right\}$ a Pauli matrix) and an entangling gate $W_i\left(\bm{\phi}_{i}\right)$. Note that we allow a parameter dependence in these entangling gates, as this will be fundamental when entangling qubits using the waveguide ansätze discussed in this work, where $\bm{\phi}_i = \left(T_i,L_i\right)$. However, these gates can still be parameter-independent, e.g., CZ or CNOT gates.

Hence, in order to successfully find the solution of a problem using a VQA, the parametrized circuit needs to generate a unitary that is as close as possible to the unitary minimizing the cost function. When no prior knowledge is known about the solution of the problem, the likelihood of such situation can be maximized using an \textit{expressive ansatz}, i.e., one that is able to explore the space of unitaries in the most complete and uniform possible way (or, equivalently, the Hilbert space of states generated as outputs of the quantum circuit). Thus, the expressibility of an ansatz (defined as the degree to which it uniformly explores the Hilbert space of the problem) is usually quantified by comparing the distribution of states generated by such quantum circuit with the uniform ensemble of Haar-random states~\cite{Sim2019ExpressibilityAlgorithms,Nakaji2021ExpressibilityComputation}
\begin{equation}
    A^t (C) = \int_{\mathrm{Haar}} \left(\ket{\psi}\bra{\psi}\right)^{\otimes t}\mathrm{d}\psi - \int_{\Theta} \left(\ket{\psi_{\theta}}\bra{\psi_{\theta}}\right)^{\otimes t}\mathrm{d}\theta\,,
\end{equation}
where $\int_{\mathrm{Haar}}$ denotes the integral over states distributed according to the Haar measure, $C$ is the parametrized quantum circuit of interest, and $\ket{\psi_\theta}$ is the state generated by the ansatz with parameter $\theta\in\Theta$. Since the ansatz with the smaller norm for $A^t (C)$ will be closer to an arbitrary state, we say that it has a higher expressibility.

However, it has been recently found that maximizing the expressibility of an ansatz is not the only feature to consider when designing a VQA. First, it was shown theoretically that if an ansatz forms a 2-design, then the gradients of its cost function vanish exponentially with the number of qubits~\cite{McClean2018}; and later works extended these results to arbitrary ansatze, showing that highly expressive quantum circuits exhibit flatter cost landscapes~\cite{Holmes2022}. This situation, known as the \emph{barren-plateau problem}, makes clear that the selection of an adequate ansatz for a VQA is a fundamental step when looking for practical applications, avoiding training issues and getting accurate final results.

\subsection{Review of the quantum critical spin models considered in this manuscript~\label{subsecSM:models}}

We first consider the XXZ model~\cite{Langari1998}
\begin{equation}\label{eqSM:HXZZ}
    H_{\mathrm{XXZ}} =\sum_{i} \left(\sigma_{\rm x}^i\sigma_{\rm x}^{i+1}+\sigma_{\rm y}^i\sigma_{\rm y}^{i+1}\right)-\Delta\sum_{i}\sigma_{\rm z}^{i}\sigma_{\rm z}^{i+1}
\end{equation}
at its $SU(2)$ symmetric ferromagnetic point $\Delta=1$. For large positive $\Delta$, the Hamiltonian is gapped and the state is ferromagnetically ordered. In between  $-1\le \Delta < 1$, the model has a critical phase describing the physics of a compactified free boson whose radius of compactification is given by $R = \sqrt{2-\frac{2}{\pi}\textrm{acos}(\Delta)}$. At $\Delta =1$  the model is gapless but the low energy excitations are magnons with a quadratic dispersion relation and thus there is no conformal symmetry.

The second model we consider is the transverse-field Ising model (TFIM) described by the Hamiltonian
\begin{equation}
    \label{eqSM:HIsingmain}
    H_{\mathrm{Ising}}=-\sum_{i}\sigma_{\rm x}^i\sigma_{\rm x}^{i+1}+g\sum_i \sigma^i_{\rm z}\,.
\end{equation}
The model presents two gapped phases, ordered and disordered, separated by a quantum critical point at $g=1$. The field theory description here is a conformal field theory (CFT) with $c=1/2$, and describes the physics of masless free fermions \cite{MussardoStatisticalPhysics}.

The last model we consider is an Ising model with LR interactions (LRTFIM), 
\begin{equation}\label{eq:H_LRI}
    H_{\text{ILR}} = -\sin{\theta}\sum_{i\neq j} \frac{1}{|i-j|^{\alpha}} \sigma_{\rm x}^{i}\sigma_{\rm x}^{j} + \cos{\theta}\sum_{i=1}^{N}  \sigma_{\rm z}^{i}\, ,
\end{equation} with $0\le \theta \le \frac{\pi}{2}$ dictating the ratio between the transverse field and the tendence of the spins to align spontaneously. By varying $\theta$, the model displays two phases, ordered and disordered, both gapped but with power-law decaying correlations and logarithmic scaling of the entropy~\cite{koffel12a}. For every choice of $\alpha$ there is a critical point separating the two phases where the system becomes gapless. Both the location of the critical point and the value of the exponents dictating the power-law decaying of the correlation functions depend on $\alpha$ ~\cite{Ruelle1968,Dyson1969,Dutta2001}. Here we focus on  three different values $\alpha=1/2,1,3$, and choose $\theta$ to be at the critical point~\cite{SupMatvarwqed} in each case.

\subsection{Review of the ans\"atze considered in this manuscript~\label{subsecSM:ansatz}}

A typical variational ansatz, as the one shown in Eq.~\eqref{eq:general_ansatz}, is formed by a combination of single-qubit operations and entangling layers. Although the possibilities are almost limitless, the combination of hardware errors and the barren-plateau problem described in the section above leads to a much narrower amount of possible quantum circuits. In what follows, we will briefly describe the fixed structure ans\"{a}tze typically used when implementing VQEs, as described in Ref.~\cite{Tilly2022} and references therein. It is worth noting, however, that it is possible to leverage the constraint of having a fixed structure ansatz, since recently ans\"{a}tze featuring an adaptative structure have been introduced. In this case, the gates are dynamically selected from an operator pool in order to maximize the gradients of the cost function, providing accurate results in quantum chemistry problems~\cite{Grimsley2019,Tang2021Qubit-ADAPT-VQE:Processor}. For this work, however, we decided to focus on fixed structure ans\"{a}tze, leaving adaptative ones for future works. 
\begin{figure}[t]
    \centering
    \includegraphics[width=\linewidth]{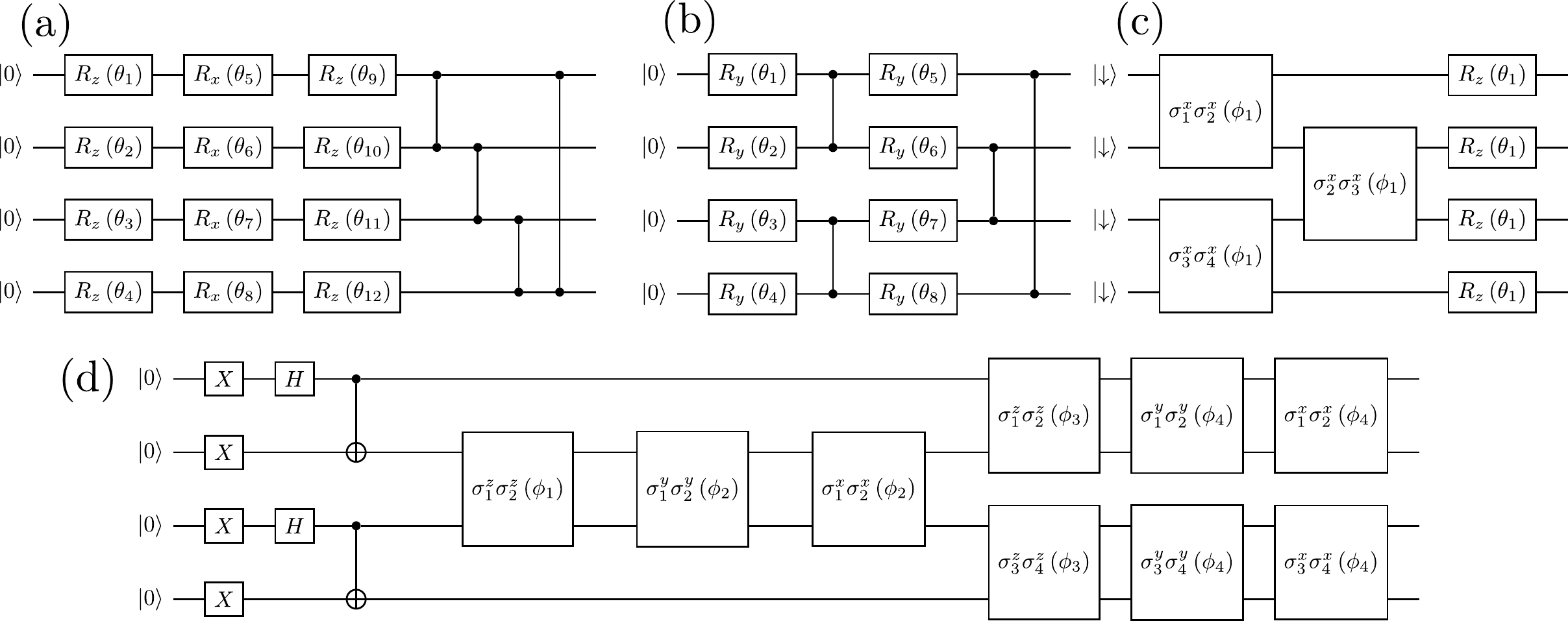}
    \caption{(a) Circuit diagram for one layer of the Hardware Efficient Ansatz (HEA) for $N=4$ qubits with a linear entanglement structure used in the main text. (b) Circuit diagram for one layer of the brick-layer ansatz, which resembles a tensor network-like structure, for $N=4$ qubits. (c) Circuit diagram for one layer of the Hamiltonian Variational Ansatz (HVA) used to calculate the ground state of the TFIM, as shown in Eq.~\eqref{eq:U_TFIM}, for $N=4$ qubits. (d) Circuit diagram for one layer of the Hamiltonian Variational Ansatz (HVA) used to calculate the ground state of the XXZ model, as shown in Eq.~\eqref{eq:U_XXZ}, for $N=4$ qubits. In this case, we had to generate a $\ket{\Phi^-}$ Bell state as initial state, and this part of the circuit was only applied once at the beggining.}
    \label{fig:ansatze_diagrams}
\end{figure}

Hence, to test the capability of our waveguide ans\"{a}tze to obtain accurate ground states for the different spin models considered, we compare with the following ans\"{a}tze:
\begin{itemize}
    \item The Hardware Efficient Ansatz (HEA), first introduced in Ref.~\cite{kandala17a} and used as the typical benchmark in other theoretical works. In its original formulation, this ansatz exploited the limited connectivity between the qubits in the underlying hardware, resulting in circuits with layers of single-qubit rotations around the $z$, $x$, and $z$ axis (to generate an arbitrary $U(1)$ rotation) followed by layers of linear entangling two-qubit gates, as shown in Fig.~\ref{fig:ansatze_diagrams}(a). The unitary corresponding to this circuit, following Eq.~\ref{eq:general_ansatz}, is
    \begin{equation}
        U_{\mathrm{HEA}}\left(\bm{\Theta}\right) = \prod_{i=1}^D \prod_{j=1}^N \mathrm{CZ}_{i,i+1} R_{i,z}^{j}\left(\theta^j_{i,z1}\right) R_{i,x}^{j}\left(\theta^j_{i,x}\right) R_{i,z}^{j}\left(\theta^j_{i,z2}\right)\, ,
    \end{equation}
    where $\mathrm{CZ}_{i,j}$ is a controlled-Z gate applied over the $i-$th control qubit and the $j-$th target qubit (with periodic boundary conditions, i.e., for a circuit with $N$ qubits one has that the $N+1$-th qubit is the same as the $1$-st one).
    \begin{figure}[b]
        \centering
        \includegraphics[width=0.3\linewidth]{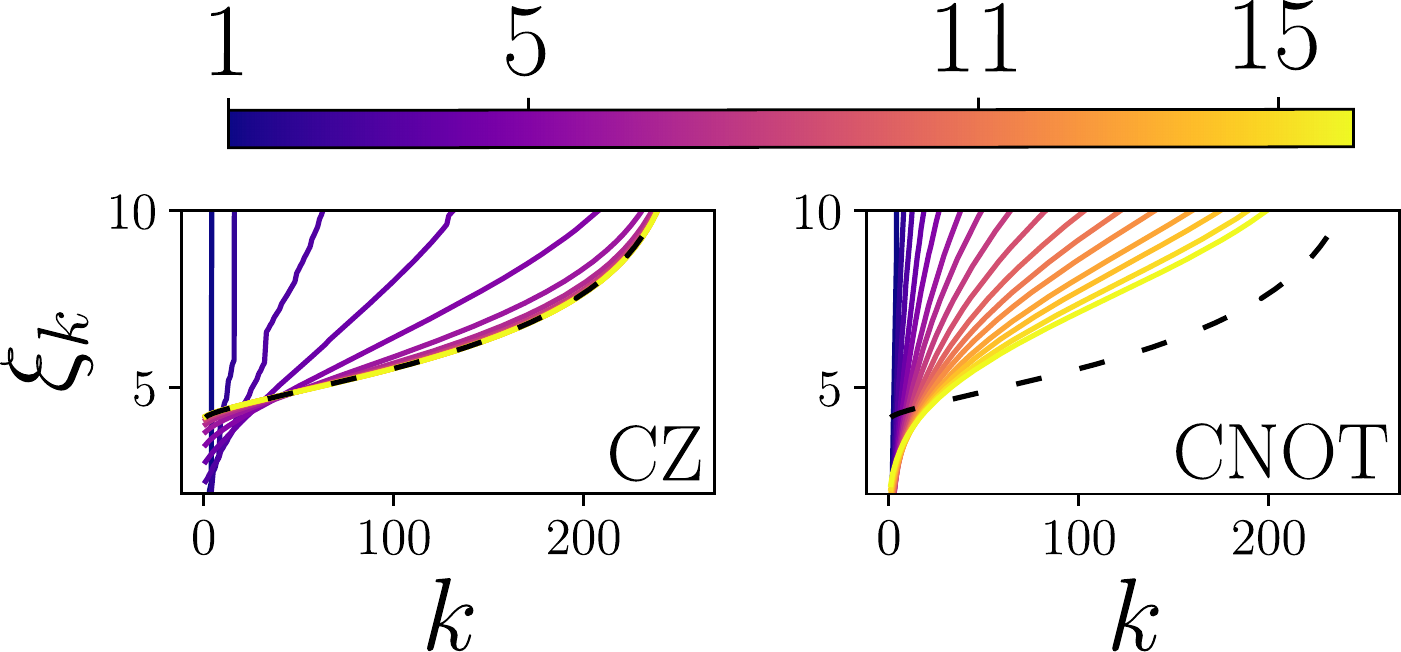}
        \caption{Entanglement spectrum of the HEA [cf. Fig~\ref{fig:ansatze_diagrams}(a)] using both CZ and CNOT gates as entangling operations for a different number of layers (in different colors). The Marchenko-Pastur distribution is shown as a black dashed line. These results were obtained considering $N=16$ qubits and calculating the mean results over $M=500$ random evaluations of the quantum circuits over their parameter space.}
        \label{fig:HEA_entanglement}
    \end{figure}

    The objective of such entangling layers is to generate the most entanglement possible. Thus, to help us decide which type of two-qubit gate use, we calculated the entanglement spectrum of two different HEAs: one using CZ gates as two-qubit gates and another one using CNOT gates instead of the CZs. These results are shown in Fig.~\ref{fig:HEA_entanglement}, where we plot the entanglement spectrum~\cite{Wiersema2020} of this ans\"{a}tze for different number of layers. Furthermore, if we consider an ensemble of random quantum states distributed according to the Haar measure (which qualitatively would mean that the whole Hilbert space of possible states is covered), the entanglement spectrum is known to follow the Marchenko-Paskur distribution of eigenvalues of random matrices, which is plotted in Fig.~\ref{fig:HEA_entanglement} as a dashed black line. Only the ansatz using CZ gates eventually saturates the Marchenko-Pastur distribution, therefore justifying its use in the main text.

    \item The brick-layer ansatz~\cite{Bravo-Prieto2020ScalingSystems} implements the entangling layers alternating between even and odd qubits, following a scheme inspired by the time-evolution block-decimation (TEBD) algorithm. Since in finite systems the gap will never close, any initial state can be adiabatically deformed to another one in the same phase under a continuous time-evolution, whose trotterization scheme under the TEBD algorithm inspires the structure of this ansatz. An example for $N=4$ qubits and one layer is shown in Fig.~\ref{fig:ansatze_diagrams}(b). In the actual simulations, we also included a final layer of single-qubit rotations (following Ref.~\cite{Bravo-Prieto2020ScalingSystems}). For such an ansatz, the corresponding unitary reads:
    \begin{equation}
        U_{\mathrm{BLA}}\left(\bm{\Theta}\right) =  \prod_{i=1}^D R_{i,y}^{0}\left(\theta^0_{i,y}\right) \prod_{j=1}^N \mathrm{CZ}_{2i,2i+1} R_{i,y}^{j}\left(\theta^j_{i,y2}\right) \mathrm{CZ}_{2i-1,2i} R_{i,y}^{j}\left(\theta^j_{i,y1}\right)\,,
    \end{equation}
    again with periodic boundary conditions, $N+1\equiv 1$.
    
    \item The Hamiltonian Variational Ansatz (HVA)~\cite{wecker15a,Reiner2019,Verdon2019,Mele2022AvoidingAnsatz,Wiersema2020}, inspired by the quantum approximate optimization algorithm (QAOA) and also adiabatic quantum computation. Its introduction was motivated by the barren-plateau problem which appeared in HEA and brick-layer ans\"{a}tze, which implies that these ans\"{a}tze would not be able to scale to a significant number of qubits due to their random-parameter initialization. These observations suggested that VQAs require \emph{problem-specific ans\"{a}tze}, tailored in such a way that the optimization landscape is as convex as possible. To achieve this, the HVA uses all the terms in the target Hamiltonian as generators of the quantum gates applied in the circuit. Namely, if we were looking for the ground state of a Hamiltonian $H = \sum_{\alpha} H_{\alpha}$, with $\left[H_{\alpha},H_{\alpha'}\right]\neq 0$ for $\alpha\neq\alpha'$, then a depth-$D$ HVA would be given by
    \begin{equation}\label{eq:HVA_ansatz}
    U_\mathrm{HVA}\left(\bm{\Theta}\right)=
    \prod_{i=1}^{D}\prod_{\alpha}e^{-i\theta_{i}^{\alpha} H_\alpha}\,,
    \end{equation}
    where one takes as the initial state $\ket{\psi_0}$ fed to the circuit the ground state of one of the Hamiltonian terms $H_{\alpha_0}$, provided that it is not the first one acting over it. Ref.~\cite{Wiersema2020} studied the HVA in depth, showing that its problem-inspired nature makes it expressive enough to obtain accurate solutions while being structured enough to allow for an efficient optimization. The main advantage of the HVA is that it allows for an effective exploration of the \emph{relevant Hilbert space}, instead of the total Hilbert space, adapting to the quantum computing regime some of the ideas that inspired variational classical methods based on tensor-networks. Note that if the depth of the circuit is big enough, Eq.~\ref{eq:HVA_ansatz} resembles a Trotterized adiabatic time-evolution of the ground state under the Hamiltonian $H=(1-s)H_{\alpha_0}+s\sum_{\alpha} H_{\alpha}$ as this is adiabatically modified from $s=0$ to $s=1$, but with time-steps $\{\theta_{i}^{\alpha}\}$ optimized classically to obtain a more accurate description of the actual ground state. 
    
    For concreteness, consider the TFIM Model given by Eq.~\eqref{eqSM:HIsingmain}. If we assume open boundary conditions and $g>0$, a depth-D HVA for this system takes the form
    \begin{equation}\label{eq:U_TFIM}
        U_{\text{HVA TFIM}} \left(\bm{\Theta}\right) = \prod_{j=1}^{D} e^{-i\theta_j H_{z}} e^{-i\phi_j H_{xx}}\,,
    \end{equation}
    where we have labeled the terms in Eq.~\eqref{eqSM:HIsingmain} as $H_{xx}=-\sum_{i=1}^{N-1}  \sigma_{i}^{x}\sigma_{i+1}^{x}$ and $H_{z}= \sum_{i=1}^{N} \sigma_{i}^{z}$. One layer of this circuit is shown in Fig.~\ref{fig:ansatze_diagrams}.

    On the other hand, the 1D XXZ Hamiltonian is described by the Hamiltonian of Eq.~\eqref{eqSM:HXZZ}.  To implement an HVA, following Ref.~\cite{Wiersema2020}, we need to divide the chain into even and odd links, so that $H_{\mathrm{XX}}=H_{\mathrm{XX}}^\mathrm{even} + H_{\mathrm{XX}}^\mathrm{odd} $, with $H_{\mathrm{XX}}^\mathrm{even} = H_{\mathrm{xx}}^\mathrm{even} + H_{\mathrm{yy}}^\mathrm{even} + H_{\mathrm{zz}}^\mathrm{even} $ and $H_{\mathrm{XX}}^\mathrm{odd} = H_{\mathrm{xx}}^\mathrm{odd} + H_{\mathrm{yy}}^\mathrm{odd} + H_{\mathrm{zz}}^\mathrm{odd} $, where we have introduced the shortcut
    \begin{equation}
        H_{\alpha\alpha}^{\mathrm{even}} = \sum_{i=1}^{N/2}\sigma_{2i-1}^\alpha \sigma_{2i}^{\alpha}\quad\text{and}\quad  H_{\alpha\alpha}^{\mathrm{odd}} = \sum_{i=1}^{N/2}\sigma_{2i}^\alpha \sigma_{2i+1}^{\alpha}\,,
    \end{equation}
    for $\alpha=x,y,z$. Since $H_{\mathrm{XX}}^{\mathrm{even}}$ commutes with $H_{\mathrm{XX}}^{\mathrm{odd}}$ and each of these terms can be implemented individually with three consecutive $\alpha\alpha$-like interactions acting over the even and odd links, respectively, a depth-$D$ HVA circuit for such Hamiltonian is
    \begin{equation}\label{eq:U_XXZ}
        U_{\text{HVA XXZ}} \left(\bm{\Theta}\right) = \prod_{j=1}^D e^{-i\theta^{\mathrm{even}}_{j,x}H_{xx}^{\mathrm{even}}}e^{-i\theta^{\mathrm{even}}_{j,y}H_{yy}^{\mathrm{even}}}e^{-i\theta^{\mathrm{even}}_{j,z}H_{zz}^{\mathrm{even}}}e^{-i\theta^{\mathrm{odd}}_{j,x}H_{xx}^{\mathrm{odd}}}e^{-i\theta^{\mathrm{odd}}_{j,y}H_{yy}^{\mathrm{odd}}}e^{-i\theta^{\mathrm{odd}}_{j,z}H_{zz}^{\mathrm{odd}}}\,.
    \end{equation}
\end{itemize}

\subsection{Optimization protocol and numerical details~\label{subsecSM:opt}}

The adiabatically-assisted VQE~\cite{Garcia-Saez2018AddressingEigensolvers} is an algorithm used to obtain accurate ground states using an iterative version of the VQE. Generally speaking, it parametrizes the target Hamiltonian used in the cost function as
\begin{equation}\label{eq:sup_results_aavqe-H}
    H_{\text{cost}}(s) = \left(1-s\right)H_0 + s H_\text{target}\,,
\end{equation}
where $H_0$ is an initial Hamiltonian with an easy-to-prepare ground state. The algorithm starts the optimization using $H(s)$ at $s=0$ as the Hamiltonian in the cost function and a state close to the eigenstate of $H_0$ as initial state. Once the optimal parameters are found, the value of $s$ is sligthly increased and the optimization is performed again, using as an initial guess the parameters obtained in the previous optimization. This procedure is repeated until the value $s=1$ is reached and the cost function minimizes the energy of the target Hamiltonian, $H_{\text{target}}$. 

The main advantage of this algorithm is that it starts close to the ground state \emph{in each optimization}, so that the classical optimizer does not need to explore undesired regions of the system's Hilbert space to find an accurate solution. However, the flatness of the cost landscape at each step is still a problem, since an accurate solution for each of the values of $s$ is needed to reach the ground state of $H_{\text{target}}$ at the end, and this can only be achieved if the classical optimizer is able to explore the Hilbert space around the previously found solution. Hence, it is interesting to combine this algorithm with some of the strategies devised to mitigate the barren plateau problem, such as problem-inspired ansatze.

Furthermore, as it is discussed in the main text, the typical interactions found in waveguide-QED setups, $H_{\text{XX/I}}$, are long-range versions of the interacting Hamiltonians in the XXZ and TFIM, respectively. These long-range Hamiltonians can be exactly mapped to their nearest-neighbors versions in the limit $L\ll1$, i.e.
\begin{equation}
    H_{\mathrm{XX}}=  J  \sum_{i\neq j} e^{-\left|i-j\right|/L} \sigma_{eg}^{i}\sigma_{ge}^{j} \simeq J_\mathrm{XX} \sum_{i} \sigma_{eg}^{i}\sigma_{ge}^{i+1}\,,
\end{equation}
if
\begin{equation}\label{eq:conditions_adiab}
    J_\mathrm{XX}= \frac{J}{2}e^{+1/L}\quad\text{and}\quad L\ll 1\,,
\end{equation}
and equivalently for $H_{\mathrm{I}}$. Hence, when using waveguide-QED interactions, we generate quantum circuits that entangle the qubits using the same kind of interactions appearing in the Hamiltonians of interest, but extended beyond nearest neighbors using such long-range connectivities. This leads to a potentially better exploration of the Hilbert space, since more states can be reached using less layers due to the possibility of entangling distant qubits with these interactions. 

With these circuits in mind, in the main text we take the following operators as the $H_0$ and $H_{\text{target}}$ appearing in Eq.~\ref{eq:sup_results_aavqe-H}: 
\begin{itemize}
    \item \emph{XXZ model}: we take $H_0 = \sum_i \sigma_{i}^{z}\sigma_{i+1}^{z}$ and $H_{\text{target}} = H_{\mathrm{XXZ}}$.     
    \item \emph{Transverse Field Ising Model}: we take $H_0 = \sum_i \sigma_{i}^{z}$ and $H_{\text{target}} = -\sum_{i} \sigma_{i}^{x}\sigma_{i+1}^{x} + \sum_i \sigma_{i}^{z}$.

    \item \emph{Long-range Transverse Field Ising Model}: in this case, we take $H_0 = \sum_i \sigma_{i}^{z}$ and $H_{\text{target}} = -\sum_{i\neq j} \sigma_{i}^{x}\sigma_{j}^{x}/\left|i-j\right|^{\alpha}$. Note that in the $\alpha\rightarrow\infty$, $s=1/2$ limit the Hamiltonian used in the cost function is the nearest-neighbors Ising Hamiltonian in the critical point. Furthermore, in this case, we calculated numerically the spectrum of $H_{\mathrm{cost}}(s)$ for different values of $s$ and stopped the iterative optimization at an $s_{\mathrm{end}}$ such as the difference between the exact energy of the ground state ($E_{\mathrm{gs}}$) and the exact energy of the first-excited state ($E_{\mathrm{1ex}}$) was $\left|E_{\mathrm{gs}}-E_{\mathrm{1ex}}\right|<1/N^2$, with $N$ the number of qubits. This provided us with a numerical estimation of the point with the smallest gap, which would recover a phase transition in the limit $N\rightarrow\infty$.
\end{itemize}

The algorithm proceeds as follows:
\begin{enumerate}
    \item It starts from a perturbative situation, where the interaction term in the cost function (proportional to $s$) is small compared with the single-qubit term (proportional to $1-s$). In particular, in our simulations, $s=0.1$ in the first step. On the other hand, the initial parameters for the waveguide gates are set in such a way that the initial state is $\ket{\psi_0}\simeq \ket{\downarrow}^{\otimes n}$, where there are corrections arising due to the entangling layers: hence, the quantum circuit implements a perturbative solution at this level, and the different number of layers can be interpreted as an increasing order in the perturbative approximation. This situation is allowed by the flexibility provided by the entangling gates, that can be ``turned-on'' continuously and hence produce a tailored solution in this case, unlike the ansatze using CZ gates (where the generation of the entanglement cannot be controlled with any parameter). All the gates initially take parameters in the interval $0.01\times\left[0,2\pi\right)$, except the length of interaction $L$ that initially is taken as $L\sim1$. With this initialization we start close to the identity, where it is known that barren plateaus can be avoided. Furthermore, numerical tests show that the initial value of $L$ is irrelevant as long as $L_{\text{initial}}=O(1)$.
    \item  A classical optimization is performed to minimize the function
    \begin{equation}
        E_s\left(\bm{\Theta}\right)=\braket{\psi\left(\bm{\Theta}\right)|H_{\text{cost}}(s)|\psi\left(\bm{\Theta}\right)}\,.
    \end{equation}
    This yields a set of parameters $\bm{\Theta}_{\mathrm{opt}}$ with energy $E_s \left(\bm{\Theta}_\mathrm{opt}\right)$.
    \item Then, the value of $s$ in the cost function is increased slightly (in particular, in steps $\Delta s = 1/(5N)$, where $N$ is the number of qubits). The previously found parameters are used as an educated guess for a new classical optimization. This step adiabatically modifies the cost function towards the critical point.
    \item Step 3 and 4 are repeated until the desired point in the phase-diagram of $H_{\text{cost}}(s)$ is reached, so the optimization provides us with an approximation of the ground state.
\end{enumerate}

Concerning the numerical details, we combined Pennylane~\cite{Arrazola2021DifferentiablePennyLane} and custom codes to simulate the quantum circuits. In particular, we found that the Lightning quantum simulator implemented in Pennylane offered the best results in terms of performance. However, since during the development of this work, Pennylane did not allow for the implementation of a global entangling operation such as the exponential of the $H_{\mathrm{XX}}$ Hamiltonian, we used a trotterized version of this matrix exponential executed with Pennylane to obtain an educated guess that we then introduced as initial parameters in a custom-coded quantum simulator that optimized the circuits using PyTorch~\cite{Paszke2019PyTorch:Library} (implementing the desired matrix exponential but with a considerable numerical overhead).

To optimize the circuits, we employed the Adam optimizer~\cite{Kingma2014Adam:Optimization} with a learning rate $\eta=0.005$ (which we found to produce the best results after hyperparameter tuning), and finished the process when the difference of cost functions for two successive iterations was smaller than $10^{-10}$ or $1000$ iterations were performed.

\subsection{Figures of merit~\label{subsecSM:fig_of_merit}}

As performance metrics to evaluate the quality of the results, we use two: First, the infidelity between the output state of the optimized quantum circuit $\ket{\psi\left(\bm{\Theta}_{\text{opt}}\right)}$ and the actual ground state obtained numerically $\ket{\psi_{\text{GS}}}$:
\begin{equation}\label{eqSM:infidelity}
    \mathcal{I} = 1 - \mathcal{F} = 1- \left|\braket{\Psi\left(\bm{\Theta}_{\text{opt}}\right)|\Psi_{\text{GS}}}\right|\,.
\end{equation}
Note that the infidelity $\mathcal{I}$ upper bounds the difference between the expectation value of any observable $\mathcal{O}$ evaluated using the variational state and the actual ground state, since letting $\mathcal{I}<\beta$, then
\begin{equation}
    \left|\langle \mathcal{O}\rangle_{\text{GS}}-\langle\mathcal{O}\rangle_{\text{var}}\right|\leq 2c \sqrt{\beta\left(1-\beta\right)}+\beta\,,
\end{equation}
where $c$ is the operator norm of $\mathcal{O}$~\cite{Wiersema2020}. For this reason, we use the infidelity as the figure of merit to benchmark the ansätze in the main text. However, we also computed the residual energy $\epsilon$~\cite{Mele2022AvoidingAnsatz}, defined as
\begin{equation}\label{eqSM:res_energy}
    \epsilon = \frac{C_{\Psi,H}\left(\bm{\Theta}_{\text{opt}}\right)-E_\text{GS}}{E_\text{max}-E_{\text{GS}}}\,,
\end{equation}
where the cost function $C_{\Psi,H}$ [cf. Eq.~\ref{eqSM:cost_fun_energy}] is evaluated at the optimal parameters $\left(\bm{\Theta}_{\text{opt}}\right)$ that minimize the expectation value of the Hamiltonian $H$, so it yields the energy of the variational state, and $E_\text{GS}$ ($E_\text{max}$) is the lowest (highest) eigenvalue of $H$. 

In Fig.~\ref{figSM:res_and_inf}(a-b) [Fig.~\ref{figSM:res_and_inf}(c-d)] we compare the values of the residual energies $\epsilon$ and the infidelities $\mathcal{I}=1-\mathcal{F}$, respectively, for the XX [TFIM] model shown in Fig.~\ref{fig:1}(a-b) [Fig.~\ref{fig:1}(c-d)] of the main text. We note that the wQED-I ansätze and the HVAs show qualitatively and quantitatively similar results when comparing the residual energies and the infidelities, meanwhile the other types of ansätze lead to states that provide accurate results for the residual energy but not for the infidelity. Since the cost function used to optimize the parameters of the variational circuit is the expectation value of $H$ over the output state obtained from the circuit, this means that in these cases the variational quantum algorithm is able to find a state with an energy similar to that of the ground state, but that can be far from the actual ground state (where this distance is quantified in terms of the infidelity between the states). On the other hand, since both the wQED-I ansätze and the HVAs use as gates the ones generated by the terms appearing in the Hamiltonian of interest, the space of generated variational states is restricted to the symmetry sector of the ground states of the Hamiltonian, so a minimization of the energy over this subspace yields an state that is close to the actual ground state. 
\begin{figure}[h]
    \centering
    \includegraphics[width=\linewidth]{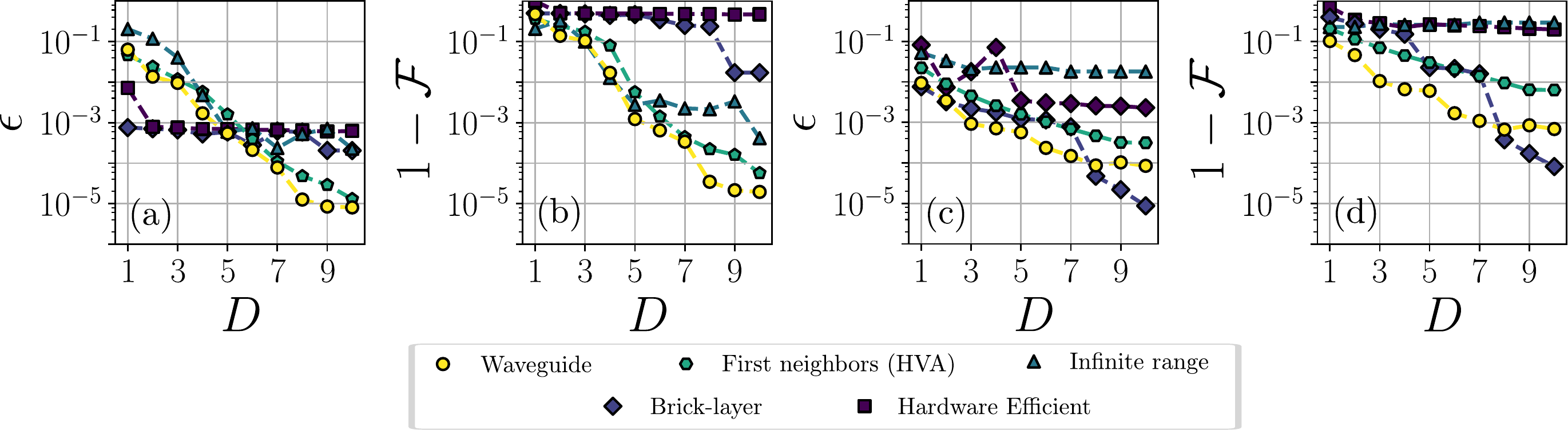}
    \caption{(a-b) Residual energy $\epsilon$ and infidelity $1-\mathcal{F}$ [Eqs.~\ref{eqSM:infidelity} and \ref{eqSM:res_energy}, respectively] obtained comparing the exact GS of the XXZ model and the optimized variational states using different ansätze as a function of the number of layers $D$ of the circuit for a system with $N=10$ qubits. (c-d) Equivalent to (a-b) but considering the TFIM with size $N=16$ qubits.}
        \label{figSM:res_and_inf}
\end{figure}

On the other hand, in Fig.~\ref{figSM:obs_for_dif_N} we plot both the residual energies $\epsilon$ and the infidelities $\mathcal{I}$ obtained when comparing the exact GS of the TFIM and the optimized variational states as a function of the number of layers of the circuits and for different sizes of the system $N$. It can be checked that the qualitative behavior (with the wQED-I ansätze and the HVAs slightly increasing their fidelities as more layers are added) is kept as the number of qubits is increased, so we only showed in the main text the results corresponding to the biggest system sizes that could be obtained with our computational resources but do not expect qualitatively different results for greater values of $N$.
\begin{figure}[h]
    \centering
    \includegraphics[width=0.9\linewidth]{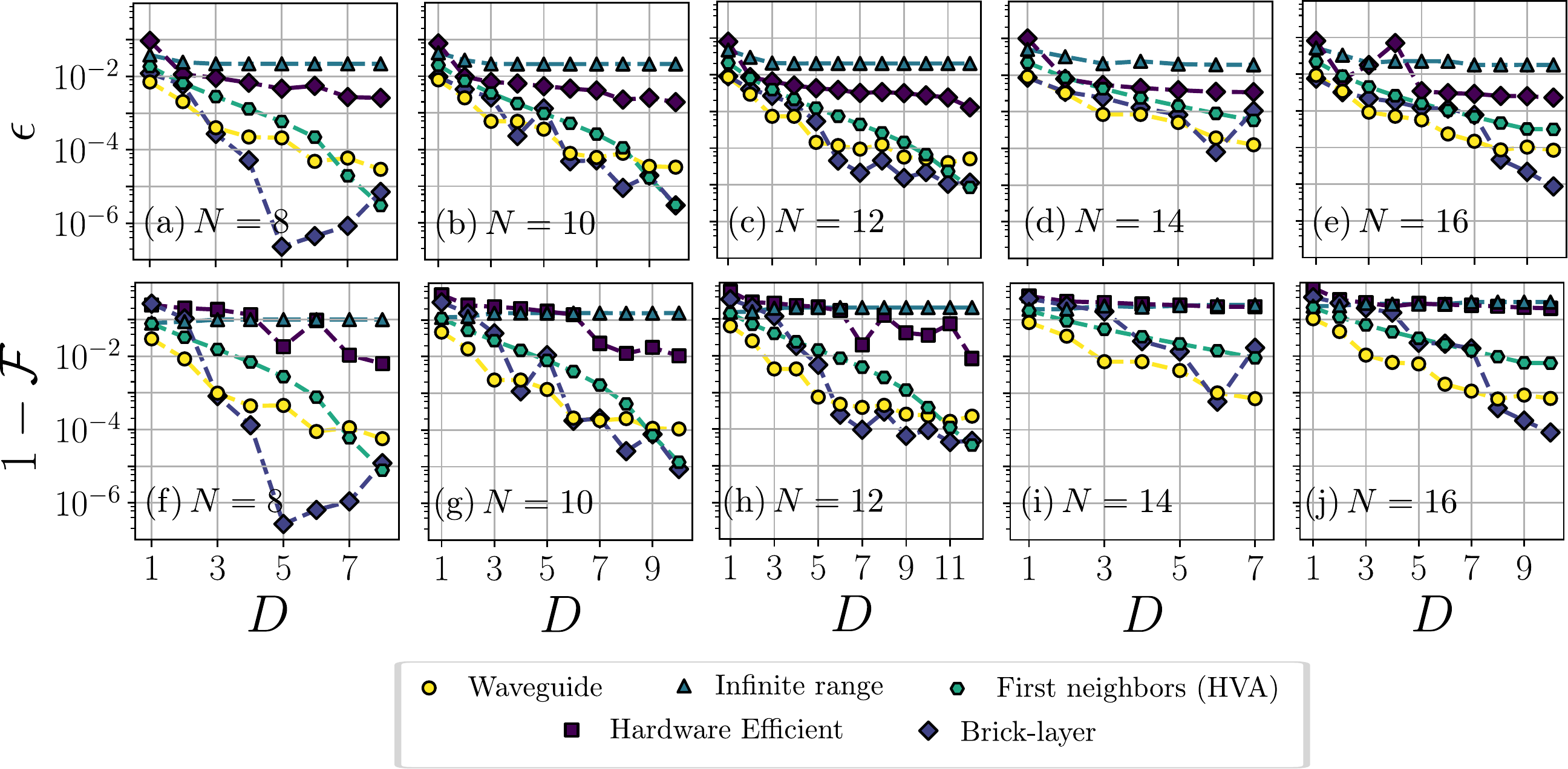}
    \caption{(a-e) Residual energies $\epsilon$ obtained by comparing the exact GS energy of the TFIM and the optimized variational ones obtained using different ansätze and circuit depths $D$, for $N=8,\,10,\,12,\,14\,\mathrm{and}\,16$ qubits, respectively. (f-j) Equivalent to (a-e) but showing the infidelities $1-\mathcal{F}$ instead of the residual energies.}
    \label{figSM:obs_for_dif_N}
\end{figure}

Furthermore, variational algorithms are intended as a way to minimize the effects of noise in quantum circuits. The consequences of this noise in the final quantum state are amplified as the depth (i.e., the number of gates in the circuit) is increased, so a good proposal for NISQ devices should minimize this depth at the same time that reaches a good approximation of the ground state. In this sense, we also quantify the quality of the different quantum circuits studied in the main text using the minimum depth needed to obtain an infidelity between the variational quantum state and the actual ground state beyond a given threshold $\delta$. That is, we reach this minimum depth $D_{\text{min}}$ if the circuit $U_{\text{min}}\left(\bm{\Theta}_{\text{opt}}\right) = \prod_{i=1}^{D_{\text{min}}} \prod_{j=1}^{N} R^{j}_{i}(\theta^{j}_{i}) W_i\left(\bm{\phi}_{i}\right)$ generates a quantum state $\ket{\Psi_\text{min}\left(\bm{\Theta}_{\text{opt}}\right)}$ such as:
\begin{equation}
    \mathcal{I} = 1 - \left|\braket{\Psi_\text{min}\left(\bm{\Theta}_{\text{opt}}\right)|\Psi_{\text{GS}}}\right|<\delta\,.
\end{equation}

\section{Error modeling~\label{secSM:error}} 

Here, we outline the general noise model used to introduce the effects of decoherence in the simulations shown in Fig.~\ref{fig:4}. As mentioned in the main text, the choice of the platform agnostic model of Ref.~\cite{kandala17a} is motivated by having a first approximation of the impact of noise for these ans\"atze in a general way without entering into the discussion of the different sources of noise that can appear in different wQED platforms, which can have a very different physical origin~\cite{goban13a,goban15a,hood16a,Samutpraphoot2020,laucht12a,evans18a,Appel2021,Tiranov2022,MacHielse2019,Rugar2020,Rugar2021,liu17a,Mirhosseini2018a,Sundaresan2019,Scigliuzzo2022,Zhang2022Simulator,krinner18a}. This model approximates the noise processes as the successive application of amplitude damping and dephasing channels acting over the system density matrix after a gate is applied. This means that the noiseless application of any unitary $U$ over a system in the state $\ket{\psi}$,
\begin{equation}
   \ket{\psi}\rightarrow U \ket{\psi}\,,  
\end{equation}
now would become the following process for the equivalent density matrix $\rho=\ket{\psi}\bra{\psi}$:
\begin{equation}
    \rho  \rightarrow \tilde{\rho}=U \rho U^\dagger \rightarrow \tilde{\rho}_a = E_0^a  \tilde{\rho} E_0^{a\,\dagger} + E_1^a  \tilde{\rho} E_1^{a\,\dagger} \rightarrow \tilde{\rho}_{a,d} = E_0^d  \tilde{\rho}_{a} E_0^{d\,\dagger} + E_1^d  \tilde{\rho}_a E_1^{d\,\dagger}\,,
\end{equation}
where the matrices $E^{a}_{i}/E^{d}_{i}$ generate the amplitude damping/dephasing channel and read
\begin{equation}\label{eq:SM_errors}
    E_{0}^{a} = \begin{pmatrix} 1 & 0 \\ 0 & \sqrt{1-p_a} \end{pmatrix},\quad E_{1}^{a} = \begin{pmatrix} 0 & \sqrt{p_a} \\ 0 & 0 \end{pmatrix},\quad E_{0}^{d} = \begin{pmatrix} 1 & 0 \\ 0 & \sqrt{1-p_d} \end{pmatrix}\quad\text{and}\quad E_{1}^{d} = \begin{pmatrix} 1 & 0 \\ 0 & \sqrt{p_d} \end{pmatrix}\,.
\end{equation}
In Eq.~\ref{eq:SM_errors}, the values $p_a$ and $p_d$ correspond to the probability of such error taking place during the computation, which depends on the specific device used. Furthermore, for simplicity, we assume that $p_a = p_d\equiv p$ (something that is generally true in superconducting devices~\cite{Barison2022VariationalComputer}), but take for the multiqubit gates probabilities of error one of order of magnitude greater, since these operations are currently the noisiest ones and hence the greatest sources of errors. Furthermore, for the wQED multi-qubit gates we assume that their error probabilities do not depend on their range. Finally, to obtain Fig.~\ref{fig:4} we first perform a noiseless adiabatically-assisted VQE to obtain the variational state $\ket{\Psi\left(\bm{\Theta}_{\mathrm{opt}}\right)}$ that approximates the ground state in the ideal case, and then use the parameters $\bm{\Theta}_{\mathrm{opt}}$ obtained as an educated guess to run the VQE only at the point of interest, but including the amplitude and phase damping channels. 

We are aware that this model does not reproduce all the complexity of actual experimental setups, and that more details should be taken into account for each specific platform. For example, the whole optimization process should consider noise, since it is known that noisy quantum circuits have flatter landscapes~\cite{Wang2021c} and are much more difficult to optimize; shot noise arising due to a finite number of measurements should be included, and also imperfect gate control. However, this simple noise model is a \emph{best-case scenario}, and even in such a situation, the fidelity of the circuits involving a greater amount of two-qubit gates (such as the brick-layer ones) perform worse than the wQED ansätze. From this analysis, we expect that more complicated error models would lead to the same conclusion: the smaller depths of the wQED ans\"atze  can be advantageous for VQA also in realistic scenarios. We leave for future work such more refined analysis.

\end{widetext}
\end{document}